\documentstyle[aps,twocolumn,floats,epsf]{revtex}
\begin{document}  
\draft
\title{Chaos, interactions, and nonequilibrium effects in the 
tunneling resonance spectra of small metallic particles}                      
         
\author{Oded Agam$^{\dag}$, Ned S.~Wingreen$^{\dag}$, 
and Boris L.~Altshuler$^{\dag \ddag}$}  
\address{$^\dag$NEC Research Institute, 4 Independence Way, Princeton,
 NJ 08540 \\
$^\ddag$Department of Physics, Princeton University, Princeton, NJ 08544}
\author{D.~C.~Ralph$^{\sharp}$ and M. Tinkham$^{\flat}$}
\address{$^{\sharp}$Laboratory of Atomic and Solid State Physics, 
Cornell University,  Ithaca, NY  14853 \\ $^{\flat}$Department of Physics
and Division of Engineering and Applied Science, Harvard University, 
Cambridge, MA 02138}
\maketitle
     
\begin{abstract}
We explain the observation of clusters in the tunneling resonance 
spectra of small metallic particles of few nanometer size.  
Each cluster of resonances is identified with one excited 
single--electron state of the metal particle, shifted 
as a result of the different nonequilibrium occupancy 
configurations of the other single--electron states. 
Assuming the underlying classical dynamics of the electrons to be 
chaotic, we determine the typical shift to be $\Delta/g$ where $g$ 
is the dimensionless conductance of the grain.
\end{abstract}     
\par                                         
\vspace{0.5cm}


An interacting many-body system exhibits, in general, a
very complicated behavior. Usually, one can analytically  
characterize only statistical properties of the spectrum. 
The fact that, for high enough energies, these properties 
are very well described by random matrix 
theory (RMT) \cite{Mehta91} was first attributed to
the complexity of the many-body system. More recently, it has 
become clear that RMT also describes single-particle 
quantum dynamics which is chaotic in the classical limit 
\cite{Casati80,Bohigas84}. Examples are non-interacting electrons
in small disordered metallic grains \cite{Efetov83}, and  
in ballistic quantum dots \cite{Andreev96}. 
Real systems, however, contain a large number of interacting 
particles, and a question which naturally arises is how does 
chaos in a single-particle description manifest itself 
in the properties of the true many-body problem?

Experimental \cite{Sivan94} as well as theoretical 
\cite{Sivan94a,Altshuler96} studies of this problem, have been 
mainly focused on two issues: the statistical properties 
of the ground state energy of quantum dots as the number of electrons changes, 
and the lifetime of a quasiparticle in such structures.
Here we consider the nonequilibrium tunneling resonance spectra  
of small metallic particles \cite{Ralph96}. 
These spectra can be measured experimentally
with high precision [see Figs.~1(a,b)] and interpreted within the 
Hartree--Fock approximation. They constitute a clear demonstration
of the interplay between many-body interactions and quantum chaos, 
and also provide direct information on the quantum chaotic 
nature of the system. 

\begin{figure}

\vspace{-0.3in}
  \begin{center}
 \leavevmode
    \epsfxsize=9cm
 \epsfbox{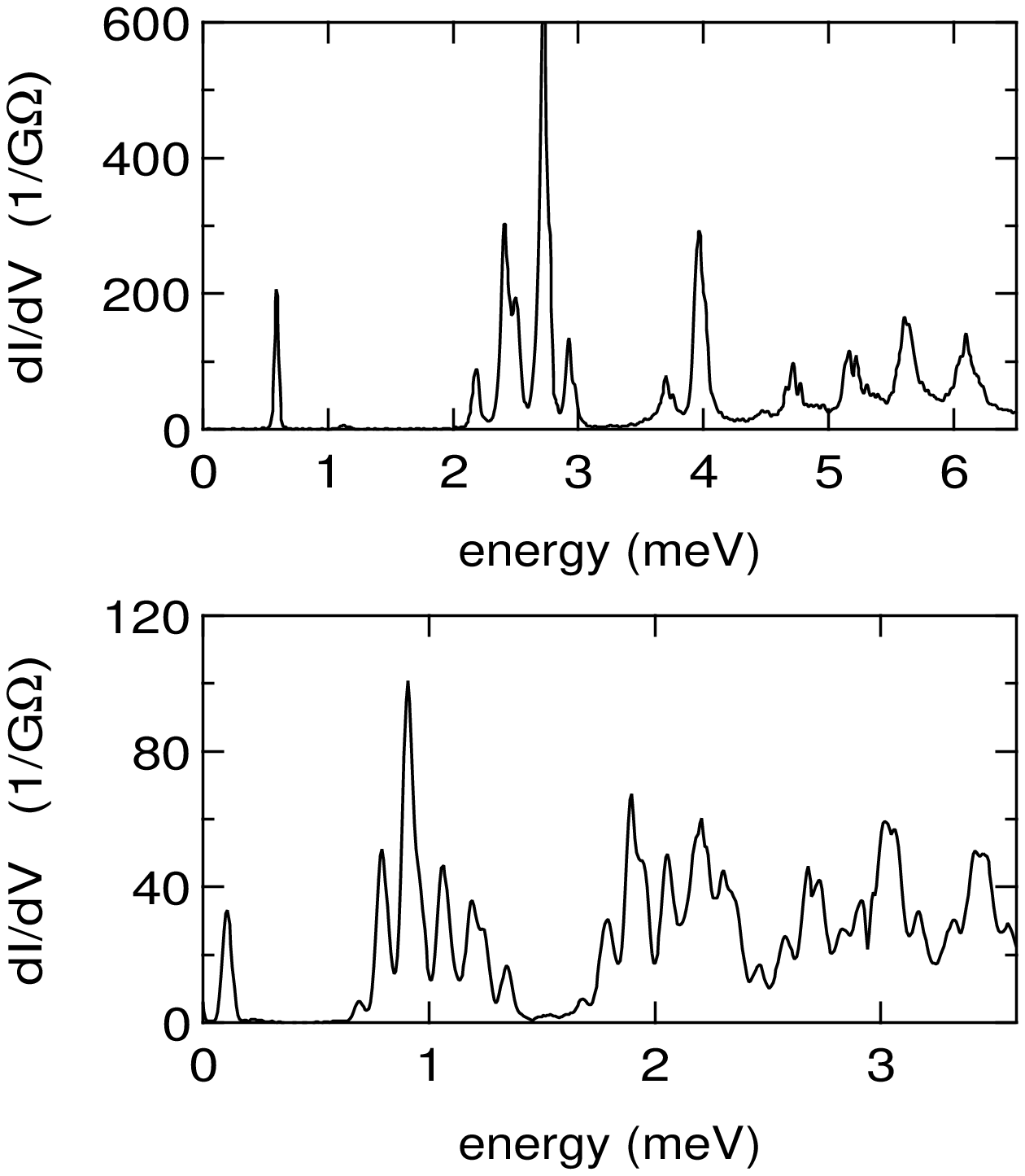} 
 \end{center}

\vspace{-3.6in}
\hspace{0.9in}
{\bf (a)}

\vspace{1.6in}
\hspace{0.9in}
{\bf (b)}

\vspace{1.2in}

  \begin{center}
 \leavevmode
    \epsfxsize=9.0cm
 \epsfbox{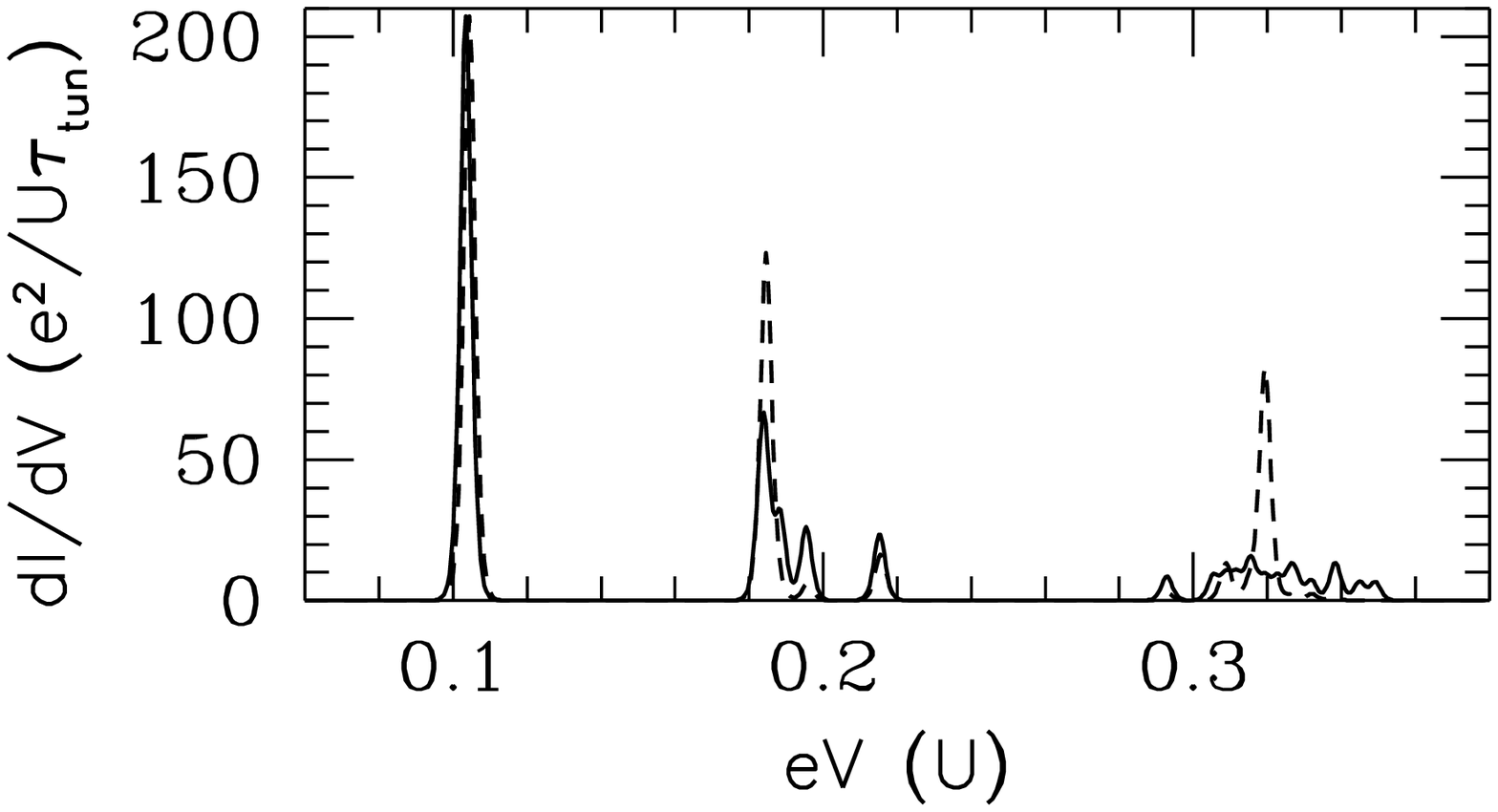}
\ 

\vspace{-3.3in}
\hspace{-0.8in}
{\bf (c)}
\vspace{1.3in}
 
\end{center}

 \caption{(a,b): The low temperature (30 mK) differential conductance
$dI/dV$ versus bias energy of small Al particles with volumes 
(a) $\approx$ 40 nm$^3$ (b) $\approx$ 100 nm$^3$ (Ref.~[9]).
The first resonance is isolated while subsequent
resonances are clustered in groups. The distance between
nearby groups of resonances is approximately the 
single--particle mean level spacing $\Delta$.
(c): Model differential conductance obtained 
from nonequilibrium detailed-balance equations: 
solid line -- in the absence of inelastic processes, 
$1/\tau_{in}=0$; dashed line -- with inelastic
relaxation rate larger than the tunneling rate, 
$1/\tau_{in}=5/\tau_{tun}$.}
  \label{fig:1}
\end{figure}

The experimental system consists of a single aluminum particle 
connected to external leads via high resistance (1 -- 5 M$\Omega$) 
tunnel junctions formed by oxidizing the surface of the particle.
In Figs.~1(a,b) we plot the differential conductance, $dI/dV$, of  
two different particles (of sizes roughly 2.5 and 4.5 nm) 
as a function of the source--drain bias energy $eV$.
The spectra display three clear features: 
(1) The low energy resonances are grouped in clusters. The distance between 
nearby clusters is of order the mean level spacing $\Delta$ of the 
noninteracting electrons in the dot. (2) The first cluster contains
only a single resonance. (3) Higher clusters consist of several 
resonances spaced much more closely than $\Delta$.

In this Letter, we explain these features as consequences of 
the underlying chaotic dynamics of the confined electrons. 
Each cluster of resonances is identified 
with one excited single-electron state, and each resonance in turn
is associated with  a different occupancy configuration of the 
metal particle's other single-electron states. The appearance of 
multiple resonances reflects the strongly nonequilibrium state of
the particle.

Our model for the system is given by the Hamiltonian:
$H=H_0+H_{T}+H_{\mbox{int}}$. Here $H_0$ describes the 
noninteracting electrons in the left (L) and right (R) 
leads and in the metallic grain,
\begin{equation}
H_0= \sum_{\alpha =L,R} \sum_q \epsilon_{\alpha q} 
d^\dagger_{\alpha q} d_{\alpha q}+ \sum_l \epsilon_l c^\dagger_l c_l.
\end{equation}
Tunneling across the barriers is described by 
\begin{equation}
H_{T}= 
\sum_{\alpha =L,R} \sum_{q,l} T^{(\alpha)}_{ql}d^\dagger_{\alpha q}c_l +
 \mbox{ H.c.},
\end{equation}
where $T^{(\alpha)}_{ql}$ are the tunneling matrix elements. Interaction
effects are taken into account only for the electrons in the grain, but 
including screening by image charges in the leads. Thus
\begin{equation}
H_{\mbox{int}}=\frac{1}{2} \sum_{ijkl}U_{ijkl} c^\dagger_i 
c^\dagger_jc_k c_l,
\end{equation}
where $U_{ijkl}$ is the matrix element of the Coulomb interaction 
for the electrons inside the grain. We remark that for the small
aluminum grains considered here one can neglect 
superconducting pairing since the single-particle mean level spacing,
$\approx \! 1$ meV, is larger than the BCS superconducting gap
which is $0.18$  meV \cite{Delft96}.   

The interaction term of the electrons is 
generally approximated by $\sum_{ijkl}U_{ijkl} c^\dagger_i c^\dagger_jc_k c_l
\approx (e\sum_l c^\dagger_l c_l )^2/C$, where
$C$ is the effective capacitance of the grain. 
Within this approximation, known as the orthodox model \cite{Averin91}, 
the charging energy depends only on the total number of electrons in 
the dot, but not on their particular occupancy configuration. 
The orthodox model is able to account for the
Coulomb blockade \cite{Averin91}, and the Coulomb staircase behavior 
of the current as the  number of extra tunneling electrons in the dot
increases. It can also be generalized to describe features
on the scale of the single-particle level spacing \cite{Averin90}.
However, the orthodox model cannot account for the clusters
of resonances in Fig.~1(a,b), since these result from fluctuations, 
$\delta U$, in the interaction energy between pairs of electrons.
Before discussing the origin of these fluctuations
we examine their effect on the differential 
conductance of the dot.

We focus our attention on the (experimental) voltage regime  
where there is no more than one extra tunneling electron in the dot. 
At small voltage bias, $V$, within the Coulomb-blockade regime 
[Fig.~2(a)], current does not flow through the system. 
Current first starts to flow when one state $i$ inside the grain
becomes available for tunneling through the left barrier, say, 
as illustrated in Fig.~2(b). As the system becomes charged 
with an additional electron, the potential energy of
the other electrons in the dot increases by $U\simeq e^2/C$, 
and some of the lower energy occupied electronic states
are raised above the right lead chemical potential [in Fig.~2(b) 
these ``ghost'' states are shown as dashed lines].
\begin{figure}
  \begin{center}
 \leavevmode
    \epsfxsize=9cm 
\epsfbox{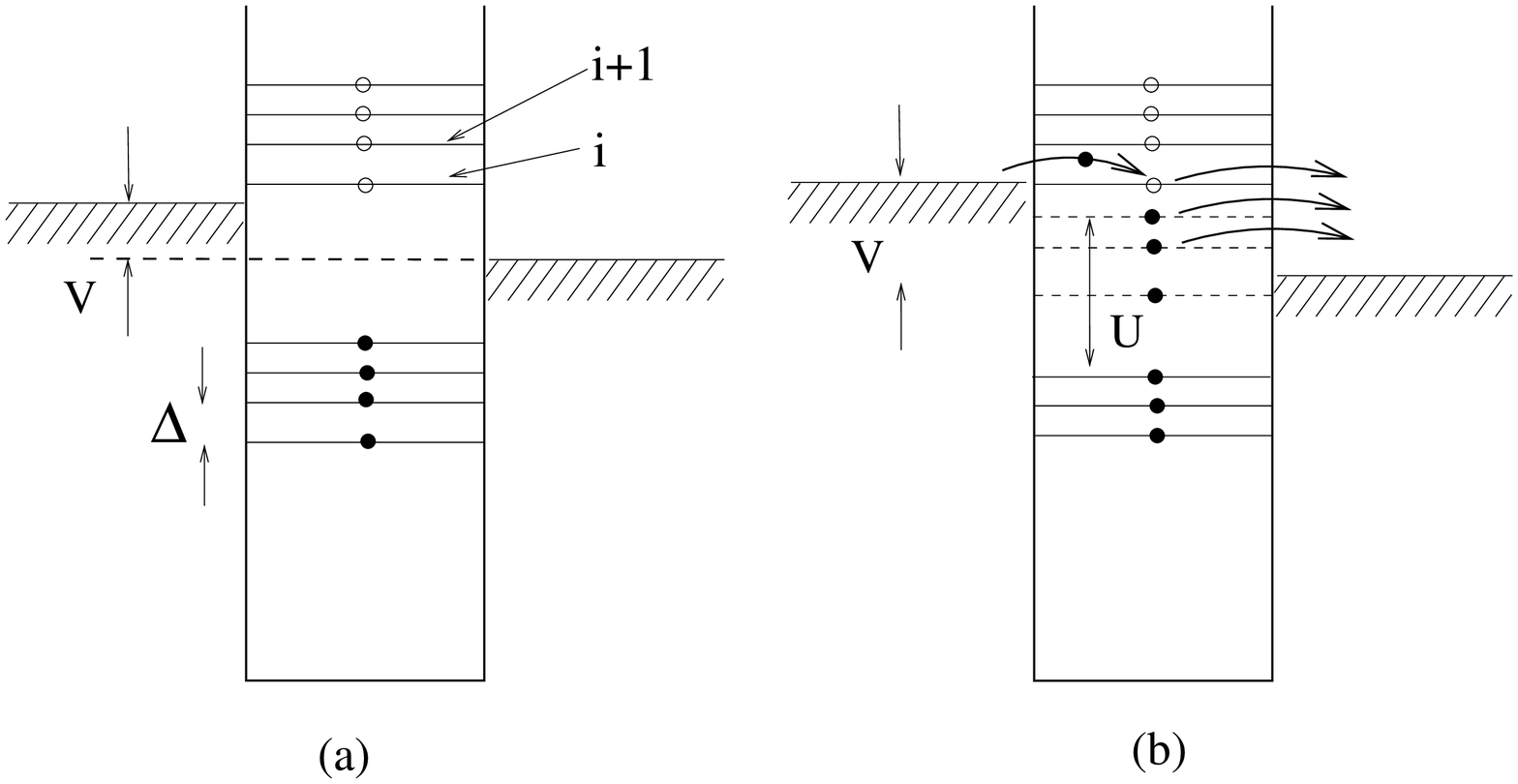}
 \end{center}
  \begin{center}
 \leavevmode
    \epsfxsize=8cm 
\epsfbox{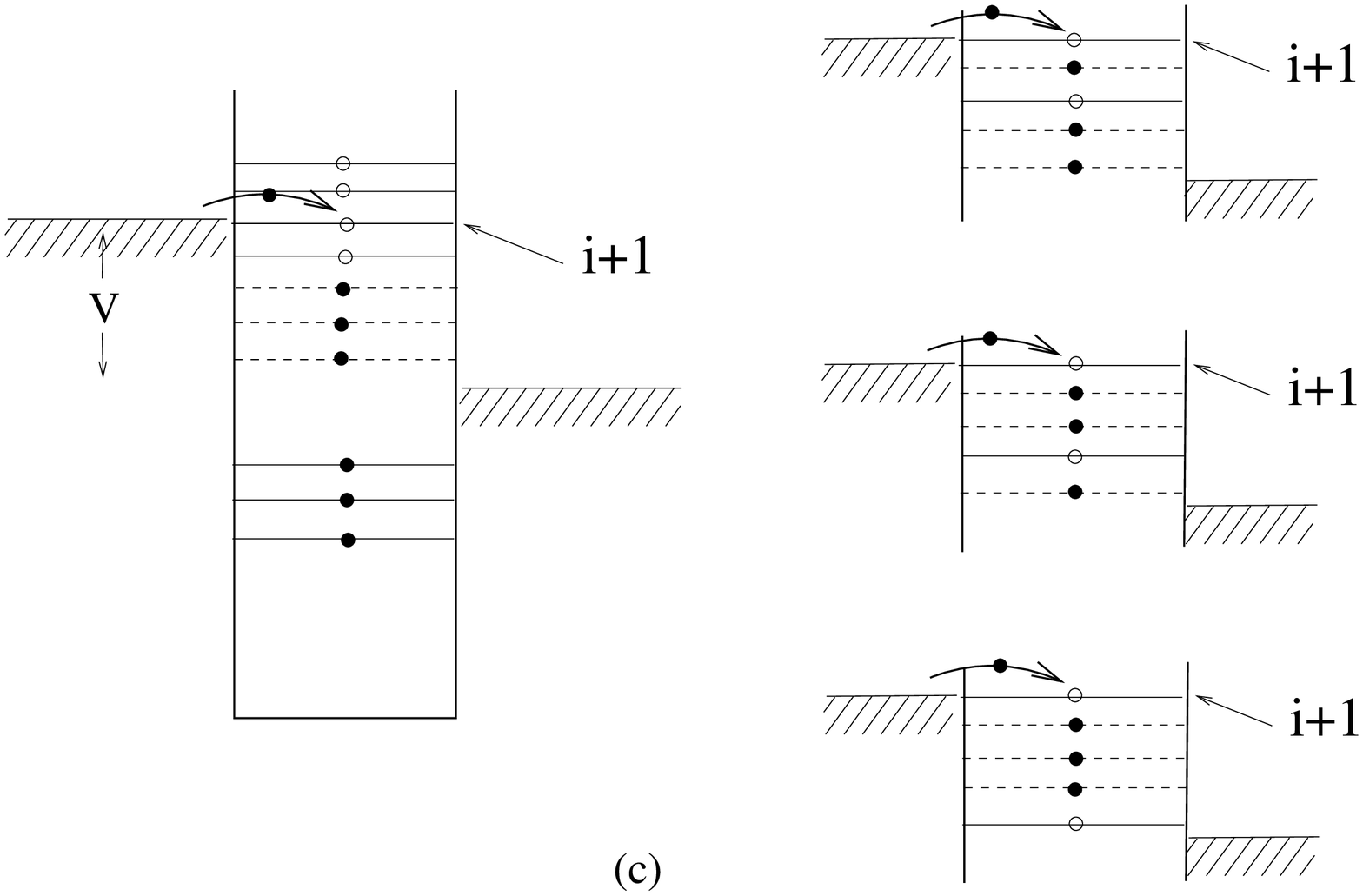}
 \end{center}
\caption{An illustration of transport through the metal particle
at various values of the source--drain voltage $V$. Filled single--particle 
levels are indicated by full circles \& empty ones by open circles.  
$U$ is the charging energy, and $\Delta$ is the single--particle 
mean level spacing. (a) The system at small voltage bias within the 
Coulomb blockade regime; (b) $V$ corresponding to the first resonance
in Figs.~1(a,b). The thin dashed lines
indicate the energy of a level after an electron has tunneled
into the dot;  (c) $V$ near
the first cluster of resonances in Figs.~1(a,b). The
splitting within the first cluster originates from the 
sensitivity of level $i+1$ to the different possible
occupation configurations as shown.}
  \label{fig:2}
\end{figure}
Electrons can tunnel out from these states into the right lead
leaving the particle in an excited state.
There is, however, only one configuration of the electrons which allows 
an electron to tunnel into level $i$ 
from the left lead, namely all lower energy 
levels occupied. This implies that only a single resonance peak appears 
in the differential conductance at the onset of the current flow 
through the system (broken spin degeneracy would cause splitting 
of this peak).

The situation changes when $V$ increases such that 
electrons can tunnel from the left lead into the next higher 
available state $i+1$, as shown in Fig.~2(c). In this case, 
there are several possible occupancy configurations, 
on which the exact energy of level $i+1$ depends. 
The several possible energies of level $i+1$ lead to  
a cluster of resonances in the differential conductance 
of the grain. The scenario described above holds provided that 
inelastic processes are too slow to maintain equilibrium in the particle.

To explicitly demonstrate the splitting of resonances induced by 
fixed fluctuations in the interaction energy $\delta U$, 
model detailed-balance equations \cite{Averin90} were solved 
numerically and the corresponding differential conductance 
plotted in Fig.~1(c) by the solid line. The model system consists of 7 equally
spaced levels, occupied alternately by 4 or 5 electrons, 
in a current-carrying steady state. For simplicity, the tunneling 
rate into each level, $1/\tau_{tun}$ ($\Gamma_{L(R)}(\epsilon_l)$ in 
the notation of Ref.~\cite{Averin90}), 
is chosen to be uniform, and the voltage is applied by increasing
the left chemical potential. The temperature is 1\% of the 
mean level spacing $\Delta$, and the variance of the 
fluctuations $\delta U$ in the interaction energy is $\Delta/5$.   
[In the absence of fluctuations ($\delta U=0$), $dI/dV$ consists of
single resonances spaced by $\Delta$.] 

To estimate the fluctuations in the interaction energy consider 
the Hartree term of the interaction energy, $U_H$. 
We wish to calculate the interaction energy difference associated 
with different occupation configurations of low energy states. 
Suppose that, as illustrated in Fig. 2(c), these  differ by a single 
occupation number, namely, in one configuration the state $j$ is empty 
and $j'$ is full while in the other $j'$ is empty and $j$ is full. 
Then 

\begin{eqnarray}
\delta U_H \!\!= \!\!\!\int \!\! dr_1 dr_2 |\psi_i({\bf r_1})|^2
U({\bf r_1},{\bf r_2})\! \left[ |\psi_{j'}({\bf r_2})|^2 \!\!-\!
|\psi_{j}({\bf r_2})|^2 \right] 
\end{eqnarray}
where the index $i$ labels an electron state other than $j$ or $j'$, 
$U({\bf r_1},{\bf r_2})$ is the interaction potential.
Clearly $\langle \delta U_H \rangle =0$, where $\langle \cdots \rangle$
denotes ensemble or energy averaging. We are therefore interested in
fluctuations of $\delta U_H$ which emerge from the non-uniform 
probability distributions of the single-particle eigenstates
in real space. To calculate $\langle \delta U_H^2 \rangle$
we approximate the interaction by $U({\bf r_1},{\bf r_2})\simeq 
\nu^{-1}\delta ({\bf r_1}-{\bf r_2})$ where $\nu$ is the density of states,
then
\begin{equation}
\langle \delta U_H^2 \rangle = 2 \nu^{-2} \int d^dr d^dr' {\cal C}^2
({\bf r,r'}), \label{Hartree-fluctuations}
\end{equation}
where ${\cal C}({\bf r, r'})=\langle |\psi ({\bf r}) \psi({\bf r'})|^2 
\rangle -\langle |\psi ({\bf r})|^2\rangle\langle |\psi ({\bf r'})|^2\rangle$ 
is the probability-density correlation function. 
For disordered systems it takes the form \cite{Gorkov65}
\begin{equation}
{\cal C}({\bf r, r'}) = \frac{\alpha \Delta}{\pi \hbar \Omega} 
\sum_{{\bf n} \neq{\bf 0}}
\frac{\phi_{\bf n}^*({\bf r})\phi_{\bf n}({\bf r'})}{D{\bf q}_{{\bf n}}^2},
\label{correlation-function}
\end{equation}
where $\alpha$ is a symmetry factor (2 for GOE systems and 1 for 
GUE), $\Omega$ is the volume of the grain, $D$ is the diffusion constant,
and the sum is over the diffusion modes $\phi_{\bf n}({\bf r})$.
Introducing the dimensionless conductance $g=\hbar \pi^2 
D/L^2 \Delta =E_c/\Delta$ where $L$ is the linear 
size of the system we obtain from
(\ref{Hartree-fluctuations}) and (\ref{correlation-function})
\cite{Blanter96} 
\begin{equation}
\langle \delta U_H^2 \rangle = \left( c \frac{\Delta}{g}\right)^2
\label{fluc}
\end{equation}
where $c = \sqrt{2} \alpha \sum_{\bf n}|{\bf n}|^{-4} /\pi $ is a 
constant of order unity. Eq.~(\ref{fluc}) also applies for 
general chaotic systems, with $g\simeq \gamma_1/\Delta$ where 
$\gamma_1$ is the first non-vanishing Perron--Frobenius eigenvalue 
\cite{Agam95}. In essence, smaller $g$ implies less uniform wave functions,
so fluctuations in the interaction energy increase as $g$ decreases. 
Experimentally we find $g \approx 5$. Unfortunately, an analytical
estimate of $g$ requires precise knowledge of the shape and disorder
of the particle which we lack \cite{comment}.

Within our approximation for the interaction potential
the Fock term, $\delta U_F$, is equal 
to $-\delta U_H$, thus apparently $\delta U_F+\delta U_H =0$. 
However, for a more realistic interaction potential 
$\delta U_F \neq -\delta U_H$, and moreover, the Fock term
 exists only for electrons with parallel spins. $\delta U_H$
is therefore the typical single--electron level splitting due to 
interaction.

More generally, when  $M$ available states below the highest
accessible energy level (including spin), are occupied by 
$M'<M$ electrons, there are ${\tiny (\! \! \begin{array}{c} 
M \\ M' \end{array}\! \!)}$ different occupancy configurations. 
The typical width of a cluster of resonances in this case is 
$W^{1/2} c\Delta/g$ where $W=\mbox{min}(M-M', M')$. The width of 
a cluster of resonances therefore {\em increases} with the 
source--drain voltage. The distance between nearby peaks of the 
cluster, on the other hand, {\em decreases} as 
$W^{1/2}/ {\tiny (\! \! \begin{array}{c} M \\ M' \end{array}\! \!)}$. 
This behavior can be seen in Fig.~1(c).

Central to our analysis is the assumption that the steady-state 
occupation configurations of the electrons in the dot are far 
from equilibrium. This condition holds when the rate  $1/\tau_{in}$ of 
inelastic relaxation processes is smaller than the tunneling 
rate of an electron into and out of the dot, $1/\tau_{tun}$. 
In the opposite limit, $1/\tau_{in} > 1/\tau_{tun}$, the system 
relaxes to equilibrium between 
tunneling events, and the electrons effectively occupy only one configuration. 
In this case one expects each resonance cluster to collapse 
to a single peak. This behavior is illustrated by the dashed line
in Fig.~1(c) where a large inelastic relaxation rate $1/\tau_{in}
=5/\tau_{tun}$ was included in the detailed-balance equations.

The results shown in Fig.~1 indicate that the metal particle in the
experimental system is indeed in a strongly nonequilibrium state. 
It is useful, however, to consider the various relaxation processes 
in our system in order to delimit the expected nonequilibrium regime.
Relaxation of excited Hartree--Fock states may occur due to: (1) 
electron-electron interaction in the dot beyond Hartree--Fock; 
(2) electron-phonon interaction; 
(3) Auger processes in which an electron in the dot relaxes 
while another one in the lead is excited; (4) relaxation of an 
electron in the dot as another electron tunnels out to the lead; (5) 
thermalization with the leads via tunneling. The last two processes 
are small corrections since they clearly happen on time scales larger than 
the tunneling time. 

In Ref.~\cite{Altshuler96} it was shown that excited  
many-body states of closed systems with energy $\epsilon$ smaller than 
$(g/\log g)^{1/2} \Delta$ are merely slightly perturbed Hartree--Fock 
states. In other words, the overlap between the true many-body 
state and the corresponding Hartree--Fock approximation is
very close to unity. This justifies the use of our model for the low
energy resonances since $g \approx 5$  therefore the energy interval 
$0\! <\! \epsilon \! < \! (g/\log g)^{1/2}\Delta$ contains at least
the first few excited states. 
At high source--drain voltage, however, when the dot is excited to
energy  $ g^{1/2}\Delta\! <\! \epsilon \! <\! g\Delta$, 
tunneling takes place into quasiparticle states of width 
$\epsilon^2/(g^2 \Delta)$ \cite{Sivan94a}. 
This width is larger than the typical separation between 
nearby resonances but smaller than $\Delta$.
Therefore, electron-electron scattering will obliterate 
the fine structure of resonances for high energy excitations of the dot.

Consider now the electron--phonon interaction. 
The temperature, 30 mK, is much smaller than the
mean level spacing, therefore, the probability of phonon
absorption is negligible, and only emission may take place. 
The sound velocity in aluminum is $v_s=6420$ m/sec, 
therefore the wavelength of a phonon associated with relaxation
of energy $\omega \sim \Delta =1$ meV is approximately 50 \AA, 
the same  as the system size. In this regime, 
we estimate the phonon emission rate to be  

\begin{equation}
\frac{1}{\tau_{e-ph}} \sim 
\left(\frac{2}{3}\epsilon_F \right)^2 \frac{\omega^3 \tau \Delta}
{2\rho \hbar^4 v_s^5},
\end{equation}
where $\epsilon_F$ is the Fermi energy (11.7 eV in Al), and $\rho$ is 
the ion mass density (2.7 g/cm$^3$ in Al). This rate is that
of a clean metal but reduced by a factor of
$\tau \Delta/\hbar$ where $\tau$ is the elastic mean free time
\cite{Reizer86}. In ballistic systems, $\tau$ is the traversal time 
across the system of an electron at the Fermi level. Assuming  
ballistic motion this factor is of order 10$^{-3}$. 
The resulting relaxation rate for $\omega=\Delta$ is therefore of order 
$1/\tau_{e-ph} \approx 10^8$ sec$^{-1}$ which is similar to the
tunneling rate $1/\tau_{tun}\approx 6 \cdot 10^8$ sec$^{-1}$ 
(corresponding to a current of $10^{-10}$ A through the particle). 
Thus, by increasing the resistance of the tunnel junctions one 
should be able to cross over to the
near-equilibrium regime shown by the dashed line in Fig.~1(c). 
  
Relaxation due to Auger process is estimated to be negligible.
Two factors reduce this rate considerably: (1) it is exponentially 
small in $w/\chi$ where $w$ is the width of the tunnel junction 
and $\chi$ is the screening length; (2) interaction between 
electrons on both sides of the 
tunnel junction can take place only within a very limited volume.    

In conclusion, we have shown that the low-voltage tunneling-resonance 
spectrum of a small metallic grain reflects a nonequilibrium 
electron configurations each of which leads to a different
energy of the single-electron level used for tunneling.
Consequently, the tunneling resonances
appear in clusters of width $\Delta/g$.
Relaxation due to electron--phonon interaction,
which becomes important for high resistance tunnel barriers, 
will collapse the clusters. This effect can be used to probe
the electron-phonon relaxation rate in nanometer size metal
particles.

The research at Harvard was supported by NSF Grant No. 
DMR-92-07956, ONR Grant No. N00014-96-1-0108, JSEP Grant No. 
N00014-89-J-1023. The research at Cornell was supported
by the Sloan Foundation, and was performed at the Cornell 
Nanofabrication Facility, funded by the NSF (Grant No. ECS-9319005), 
Cornell University, and industrial affiliates.


\begin{thebibliography}{99}
\bibitem{comment} A naive estimate of $g$ in ballistic systems is
$\hbar/\tau \Delta$, where $\tau$ is the time for an electron
at the Fermi energy to cross the system. The metallic 
grains of the experiment, however,  have a roughly pancake shape.
Assuming diffusive dynamics one can show that $g\tau\Delta/\hbar 
\propto (z/r)^2$ where $z$ is the pancake thickness and $r$ is its radius.
$g$ is therefore much smaller than $\hbar/\tau \Delta$.
\end{thebibliography}
\end{document}